\newcommand{\be}{\begin{equation}}
\newcommand{\ee}{\end{equation}}
\newcommand{\bea}{\begin{eqnarray}}
\newcommand{\eea}{\end{eqnarray}}
\newcommand{\nn}{\nonumber}
\newcommand{\p}[1]{(\ref{#1})}
\newcommand\T{\theta_{12}}
\newcommand\Tb{{\bar\theta}_{12}}
\newcommand\Z{Z_{12}}
\newcommand\D{{\cal D}}
\newcommand\Db{\overline{\cal D}}

\documentstyle[12pt]{article}
\begin{document}
\renewcommand{\thefootnote}{\fnsymbol{footnote}}
\thispagestyle{empty}
\begin{flushright}
JINR E2-95-538\\
hep-th/9512214
\end{flushright}
\vspace{2.5cm}
\begin{center}
QUANTUM $N=2$ SUPER $W_3^{(2)}$ ALGEBRA IN SUPERSPACE
\vspace{1cm} \\
{\sc Changhyun Ahn}$^{1)}$\footnote{E-mail: cahn@photon.kyunghee.ac.kr},
{\sc E. Ivanov}$^{2)}$\footnote{E-mail: eivanov@thsun1.jinr.dubna.su},
{\sc S. Krivonos}$^{2)}$\footnote{E-mail: krivonos@thsun1.jinr.dubna.su}
and
{\sc A. Sorin}$^{2)}$\footnote{E-mail: sorin@thsun1.jinr.dubna.su}
\vspace{1cm}\\
$^{1)}$ {\it Department of Physics, Kyung Hee University, \\
Seoul 130-701, Korea}\vspace{0.5cm}\\
$^{2)}${\it Bogoliubov Laboratory of Theoretical Physics, JINR, \\
141980 Dubna, Moscow Region, Russia}
\vspace{3cm} \\
{\sc Abstract}
\end{center}

We discuss the $N=2$ extension of Polyakov-Bershadsky
$W_3^{(2)}$ algebra with the generic central charge, $c$, at the
{\it quantum} level in superspace. It contains, in addition to the
spin $1$ $N=2$ stress tensor, the spins $1/2, 2$ bosonic
and spins $1/2, 2$ fermionic supercurrents satisfying the first class
{\it nonlinear} chiral constraints. In the $c \rightarrow \infty $ limit,
the ``classical'' $N=2$ $W_3^{(2)}$ algebra is recovered.
\vspace{1cm}
\begin{center}
December 1995
\end{center}
\vfill
\setcounter{page}0
\renewcommand{\thefootnote}{\arabic{footnote}}
\setcounter{footnote}0
\newpage

\noindent{\bf 1. Introduction}
\vspace{0.2cm}

\noindent In recent years various (super)extensions of nonlinear
$W$ algebras have been studied in two dimensional rational
conformal field theory \cite{s}.
There exist very special algebras,
quasi-superconformal algebras, which include
the bosonic currents with {\it noncanonical} half-integer spins.
One of the examples of such an algebra is
the Polyakov-Bershadsky $W_3^{(2)}$ algebra \cite{p,b}.
It is a bosonic analog of the linear
$N=2$ superconformal algebra (SCA) \cite{x}: it contains two bosonic
currents with {\it noncanonical} spins $3/2$,
besides two bosonic currents with canonical spins $1$  and $2$.
Quadratic nonlinearity in the operator product expansion (OPE) of
the spin $3/2$ currents in this algebra emerges from requiring
the associativity.

In ref. \cite{KS}, an $N=2$ supersymmetric extension of this algebra
has been constructed using the Polyakov's soldering procedure
at the classical level. This extension comprises four {\it additional}
fermionic currents with non-canonical integer spins $ (1,1,2,2)$, besides
$N=2$ SCA and $W_3^{(2)}$ algebra as two different subalgebras.
For the quantum case, it was further studied in \cite{AKS}.
A new feature of the quantum case is the necessity
to include, in the r.h.s. of OPEs, {\it extra}
composite currents which were not present in
the classical case \cite{KS}. These are essential in order that OPEs
between the twelve $N=2$ $W_3^{(2)}$ currents form a closed set,
in other words, satisfy the Jacobi identities. The currents of this
$N=2$ super $W_3^{(2)}$ algebra, both at the classical and quantum levels,
cannot be arranged into $N=2$ supermultiplets with respect to the
original (manifest) $N=2$ SCA because the numbers of currents with integer
and half-integer spins do not match each other.

Recently it has been shown \cite{IKS}, at the classical
level, that a $N=2$ superfield formulation of this algebra can be
achieved by exploring another, {\it hidden} $N=2$ SCA.
In \cite{IKS}, $N=2$ $W_3^{(2)}$ algebra has been reformulated
in terms of the spins $ (1/2, 2) $ bosonic supercurrents and
spins $ (1/2, 2) $ fermionic ones satisfying the first class
nonlinear chiral constraints, and a {\it modified} $N=2$ spin $1$
stress tensor $J(Z)$ (see the next section for notation). In \cite{ais}
this $N=2$ superfield formulation of classical $N=2$ $W_3^{(2)}$ algebra
has been reproduced in the framework of $N=2$ superfield Hamiltonian
reduction, starting from $N=2$ super affine extension of the superalgebra
$sl(3|2)$.

In the present article we show that the quantum $N=2$ super $W_3^{(2)}$
algebra of ref. \cite{AKS} also admits a similar superfield description
with respect to the {\it modified} $N=2$ SCA. We give
the relevant SOPEs in the explicit form and compare them with the
classical expressions of ref. \cite{IKS}.
\vspace{0.4cm}

\noindent{\bf 2. $N=2$ superfield structure of quantum
$N=2$  $W_3^{(2)}$ algebra}
\vspace{0.2cm}

\noindent We start by recalling the basic points of ref. \cite{IKS}
that is closely related to our study.

$N=2$ $W_3^{(2)}$ algebra in its origianl form \cite{KS}
is generated by six bosonic currents
$\left\{ J_w, J_s,  \right.$ $\left.G^{+}, G^{-}, T_w, T_s \right\}$
and six fermionic ones
$\left\{ S_1, \bar{S}_1, S, \bar{S} \right.$, $ \left.
S_2,\bar{S}_2 \right\}$
with the spins $\left\{ 1, 1, 3/2, 3/2, \right.$ $\left. 2, 2 \right\}$,
respectively. A sum of $T_w, T_s$ and some appropriate composite currents
is chosen as the Virasoro stress tensor.

In order to equalize the numbers of the half-integer and integer
spins, the authors of \cite{IKS} passed to another, twisted Virasoro
stress tensor
by making use of the fact that $\{ G^{+}, G^{-}, S_1, \bar{S}_1, S,
\bar{S},$ $S_2, \bar{S}_2 \}$ have non-zero $u(1)$ charges associated
with the spin $1$ currents $J_w$ and $J_s$. This properly twisted stress
tensor is given by the expression:
\be
T_s+T_w+\frac{4}{c}S_1\bar{S}_1-\frac{4}{c}J_s^2+\frac{12}{c}J_wJ_s-
  \frac{12}{c}J_w^2 -\partial J_s.
\ee
With respect to it the above eight currents possess, respectively, the
following spins
\be
(1/2, 5/2, 1/2, 3/2, 2, 1, 3/2, 5/2)\; .
\ee
The spins of the remaining currents $\{ J_s, J_w, T_s, T_w \}$ are
the same as before, that is, $(1, 1, 2, 2)$. This way one succeeds in
getting equal number of currents with integer and half-integer
spins\footnote{It is to the point here to remark that the common belief
that the spin $1/2$ currents can be factored out to yield a smaller
nonlinear algebra \cite{GS} is not true in the case under consideration
\cite{IKS}. The reason is that OPEs between spin $1/2$ currents
do not contain central terms the presence of which is crucial for
these currents to be decoupled \cite{GS}. By the same reason one cannot
decouple the spin $1/2$ supercurrents in the superfield form of $N=2$
super $W_3^{(2)}$.}.

Then, by adding two spins $(5/2,3)$ composite bosonic currents
and two fermionic ones of the same spins, the $N=2$ $W_3^{(2)}$ algebra
at the classical level can be rewritten in terms of the following five
$N=2$
supercurrents: a general spin 1 supercurrent $J(Z)$, spin 1/2 anti-chiral
fermionic and bosonic supercurents $G(Z)$ and $Q(Z)$,
general spin 2 fermionic $F(Z)$ and bosonic $T(Z)$ ones\footnote{
By $Z$ we denote the coordinates of $1D$ $N=2$ superspace,
$Z=(z, \theta, \tilde{\theta})$.}. To ensure the irreducible current
content of the algebra, the latter two supercurrents should be subject
to nonlinear chirality constraints \cite{IKS} the precise form of which for
the quantum case will be given below. It can be found in \cite{IKS}
how the $N=2$ $W_3^{(2)}$ currents are spread over these five
superfields. For what follows it will be important to remember that
one indepenent combinaton of the original two $u(1)$ currents $J_w$ and
$J_s$, $\tilde{J}_s$, appears as the lowest component of the $N=2$ stress
tensor $J(Z)$, while another, $\tilde J$, as the second,
highest component in the anti-chiral superfield $G(Z)$.

 From now on, we want to extend the consideration of
ref. \cite{IKS} to the quantum case. The basic new feature is that now
we should take care of Jacobi identities to all orders in contractions
between the composite supercurrents.

As a starting point we take a natural assumption that the
supercurrent $J(Z)$ generates the standard linear $N=2$ SCA
\footnote{Hereafter we do not write down the regular parts of SOPEs.
All the supercurrents appearing in the right-hand sides of the SOPEs
are evaluated at the point $Z_2$. Multiple composite currents are always
regularized from the right to the left.}
\begin{equation}
J(Z_1)J(Z_2)=-\frac{1}{\Z^2}2c+ \frac{\T \Tb}{\Z^2} J
+\frac{\Tb}{\Z} \Db J-\frac{\T}{\Z} \D J
+\frac{\T\Tb}{\Z} \partial J \; ,
\label{eq:jj}
\end{equation}
where
\begin{equation}
\T=\theta_1-\theta_2 \quad , \quad \Tb=\bar\theta_1-\bar\theta_2 \quad ,
 \quad \Z=z_1-z_2+\frac{1}{2}\left( \theta_1\bar\theta_2
-\theta_2\bar\theta_1 \right) \quad ,
\end{equation}
and $\D,\Db$ are the spinor covariant derivatives defined by
\begin{equation}
\D=\frac{\partial}{\partial\theta}
 -\frac{1}{2}\bar\theta\frac{\partial}{\partial z} \; , \quad
\Db=\frac{\partial}{\partial\bar\theta}
 -\frac{1}{2}\theta\frac{\partial}{\partial z}\; ,
\end{equation}
$$
\left\{\D,\Db \right\}= -\frac{\partial}{\partial z} \quad , \quad
\left\{\D,\D \right\} = \left\{\Db,\Db \right\}= 0.
$$

Let us also assume that all the SOPEs which in the classical case
involve no nonlinearities retain their structure in the quantum case too
(an analogous assumption in the component approach proved to be
true \cite{AKS}). Then the remaining four supercurrents have the
above-mentioned spins with respect to this $N=2$ SCA
\begin{eqnarray}
J(Z_1)G(Z_2) & = & -\frac{\T}{\Z^2} c +\frac{\T \Tb}{\Z^2}\frac{1}{2} G
 -\frac{1}{\Z} G -\frac{\T}{\Z} \D G +\frac{\T \Tb}{\Z} \partial G
\; , \nn \\
J(Z_1)Q(Z_2) & = & \frac{\T \Tb}{\Z^2} \frac{1}{2} Q
 -\frac{1}{\Z} Q-\frac{\T}{\Z} \D Q +\frac{\T \Tb}{\Z} \partial Q
\; , \nn \\
J(Z_1)F(Z_2) & = & \frac{\T \Tb}{\Z^2} 2 F
 +\frac{\Tb}{\Z} \Db F -\frac{\T}{\Z} \D F +\frac{\T \Tb}{\Z} \partial F
\; , \nn \\
J(Z_1)T(Z_2) & = & \frac{\T \Tb}{\Z^2} 2 T
 +\frac{\Tb}{\Z} \Db T -\frac{\T}{\Z} \D T +\frac{\T \Tb}{\Z} \partial T.
\label{eq:pri}
\end{eqnarray}
Note that the $u(1)$ charges $-1$ of two anti-chiral supercurrents
$G(Z)$ and $Q(Z)$ with respect to the
$u(1)$ current $\tilde{J}_s$ are strictly fixed by their
chirality property
\begin{equation} \label{chir}
\Db G(Z) =  \Db Q(Z) = 0\;.
\end{equation}
While acting on both sides of the relevant
SOPEs at the point $Z_2$, $\Db$ should yield zero.

As a next step, we are led to consider the
most general ansatz for the remaining SOPEs, such that it is consistent
with symmetry under the interchange $Z_1 \leftrightarrow Z_2$, statistics,
spins and the conservation of two $u(1)$ charges (the property that
$Q(Z)$ and $F(Z)$ have nonzero $u(1)$ charges with respect to the
$u(1)$ current $\tilde J$ strictly fixes the SOPEs of these
superfields with $\D G$ and hence with $G$ itself, see first two equations
in (\ref{eq:alg}) ). After inclusion of all possible composite currents
with undetermined structure constants
we are left with more than $200$ terms, in contrast to the classical
consideration. In the component approach \cite{AKS},
we have experienced that the quantum Jacobi identities are not satisfied
if one specializes only to those algebraic structures which are present at
the classical level. Nonetheless, we will see that only {\it three}
extra composite currents finally survive.
Our approach to fixing the structure constants is extremely direct
and straightforward: it goes by exploiting the Jacobi identities between
the supercurrents.

As a final result we arrive at the following SOPEs which obey all the
Jacobi identities except for $( T, T, T )$
\footnote{ We have not been able to check the Jacobi
identity of $( T, T, T ) $ directly in package SOPEN2defs \cite{ttt}
using the desk-top. This check can be done in an indirect way,
as discussed below.} for the generic value of the central charge
\begin{eqnarray}
G(Z_1)Q(Z_2)&=& -\frac{\T}{\Z} \frac{Q}{2}  \; , \nn \\
G(Z_1)F(Z_2)&=& \frac{\T}{\Z} \frac{F}{2} \; , \nn \\
G(Z_1)T(Z_2)&=&
-\frac{\T}{\Z^3} 2c - \frac{\T \Tb}{\Z^3} 2 G
   -\frac{1}{\Z^2} 2 G + \frac{\T}{\Z^2} \left[ J + 2 \D G \right] \nn \\
            & & +
    \frac{\T \Tb}{\Z^2} \frac{1}{(-1+2c)} \left[ \frac{1}{2}(1 + 2c) \Db J +
     4 G \D G +
    2 J G -
    (-1+2c) \partial G \right] \nn \\
            & & +
   \frac{1}{\Z} \frac{1}{(-1+2c)} \left[ (1 + 2c) \Db J + 8 G \D G  +
     4 J G - 2(-1+2c) \partial G \right] \; , \nn \\
Q(Z_1)F(Z_2)&=&
\frac{\T}{\Z^3} 2c + \frac{\T \Tb}{\Z^3} 2 G+
   \frac{1}{\Z^2} 2 G - \frac{\T}{\Z^2} \left[ J + 2 \D G \right] \nn \\
            & & +
    \frac{\T \Tb}{\Z^2} \frac{1}{(-1+2c)} \left[ -\frac{(1 + 2c)}{2} \Db J-
    4 G \D G -
    2 J G +
    (-1+2c) \partial G \right] \nn \\
            & & +
   \frac{1}{\Z} \frac{1}{(-1+2c)} \left[ (-1 - 2c) \Db J -
 8 G \D G -
     4 J G +  2(-1+2c) \partial G \right] \nn \\
            & & +
 \frac{\T}{\Z} \frac{1}{2}  T \; , \nn \\
Q(Z_1)T(Z_2)&=&
-\frac{\T \Tb}{\Z^3} 2 Q
   -\frac{1}{\Z^2} 2 Q + \frac{\T}{\Z^2} 2 \D Q \nn \\
            & & +
    \frac{\T \Tb}{\Z^2} \frac{1}{(-1+2c)} \left[ 4 G \D Q +
    2 J Q +
    4 \D G Q +
    (3 - 2c) \partial Q \right] \nn \\
            & & +
   \frac{1}{\Z} \frac{1}{(-1+2c)} \left[ 8 G \D Q + 4
 J Q +
    8 \D G Q +
    2(3 - 2c) \partial Q \right] \; , \nn \\
F(Z_1)T(Z_2)&=&
\frac{1}{\Z^2} 4 F - \frac{\Tb}{\Z^2} \frac{8}{(1 + 2c)} G F +
\frac{\T}{\Z^2}
 2 \D F  +
    \frac{\T \Tb}{\Z^2} \frac{1}{(-1+2c)} \left[ \frac{16}{(1 + 2c)} G \D F
 \right. \nn \\
           & & \left. + 4 J F +
     \frac{8(1 + 6c)}{(1 + 2c)} \D G F +
    (-1 - 6c) \partial F \right]  +
    \frac{1}{\Z} 2 \partial F  \nn \\
           & &
 +\frac{\Tb}{\Z} \frac{1}{(-1+2c)} \left[-\frac{32}{(1 + 2c)} G \D G F -
    \frac{16}{(1 + 2c)} J G F -
    4 \Db J F -
    8 \partial G F \right] \nn \\
           & & +
    \frac{\T}{\Z} \frac{1}{(-1+2c)} \left[ 2(1 + 2c) \partial \D F
 -
    4 J \D F -
    8 \D G \D F +
    4 \D J F \right] \nn \\
            & & +
    \frac{\T \Tb}{\Z} \frac{1}{(-1+2c)} \left[ 4 J \partial F -
    4 \Db J \D F +
    16 \D G \partial F  +
    \frac{16}{(1 + 2c)} \D J G F \right. \nn \\
            & & \left. +
    16 \partial \D G F -
    2(1 + 2c) \partial^2 F \right] \; , \nn \\
T(Z_1)T(Z_2)&=&
\frac{1}{\Z^2} 4 T - \frac{\Tb}{\Z^2} \frac{8}{(1 + 2c)} \left[ G T +
Q F \right] + \frac{\T}{\Z^2} 2 \D T \nn \\
            & & +
    \frac{\T \Tb}{\Z^2} \frac{1}{(-1+2c)} \left [ \frac{16}{(1 + 2c)} G \D T
    + 4 J T -
    \frac{16}{(1 + 2c)} Q \D F \right. \nn \\
            & & \left. +
    \frac{8(1 + 6c)}{(1 + 2c)} \D G T +
    \frac{8(1 + 6c)}{(1 + 2c)} \D Q F +
    (-1 - 6c) \partial T \right]
             + \frac{1}{\Z} 2 \partial T \nn \\
            & &
 +\frac{\Tb}{\Z} \frac{1}{(-1+2c)} \left[-\frac{32}{(1 + 2c)} G \D G T -
    \frac{32}{(1 + 2c)} G \D Q F -
    \frac{16}{(1 + 2c)} J G T  \right. \nn \\
            & & \left. -
    \frac{16}{(1 + 2c)} J Q F -
    4 \Db J T -
    \frac{32}{(1 + 2c)} \D G Q F
            -
    8 \partial G T -
    \frac{8(3 + 2c)}{(1 + 2c)} \partial Q F \right] \nn \\
            & & +
    \frac{\T}{\Z} \frac{1}{(-1+2c)} \left[ 2(1 + 2c) \partial \D T -
    4 J \D T -
    8 \D G \D T
             +
    4 \D J T +
    8 \D Q \D F \right] \nn \\
            & & +
    \frac{\T \Tb}{\Z} \frac{1}{(-1+2c)} \left[ 4 J \partial T
            -
    4 \Db J \D T +
    16 \D G \partial T +
    \frac{16}{(1 + 2c)} \D J G T
            \right. \nn \\
            & & \left. +
    \frac{16}{(1 + 2c)} \D J Q F
            +
    16 \D Q \partial F +
    16 \partial \D G T
             +
    16 \partial \D Q F -
    2(1 + 2c) \partial^2 T \right]\;.
\label{eq:alg}
\end{eqnarray}
Thus we have the complete structure of $N=2$ quantum $W_3^{(2)}$
algebra, (\ref{eq:jj}), (\ref{eq:pri}), (\ref{eq:alg}), in $N=2$
superspace.

Let us make several comments on this result.

The above SOPEs are consistent only on the shell of the constraints:
\begin{eqnarray}
A_1 & \equiv & \Db F +\frac{4}{1+2c} (G F) =0 \quad , \label{con1} \\
A_2 & \equiv & \Db T +\frac{4}{1+2c} (G T) +\frac{4}{1+2c} (QF) =0
\; ,
\label{eq:con2}
\end{eqnarray}
which are the proper quantum version of the constraints of the
classical case \cite{{IKS},{ais}}.
We systematically used the above two constraints while fixing the
structure constants in the SOPEs.
We have checked that these constraints
are first class as in the classical case, that is, the SOPEs
between $A_1$ and $A_2$ (as well as their SOPEs with all supercurrents)
vanish on the shell of constraints. The classical form of $A_1, A_2$
is recovered in the $c \rightarrow \infty$ limit (properly implemented).
Note that all the above SOPEs are by construction consistent with the
linear anti-chirality conditions \p{chir}.

The following correspondence between the component currents and
superfields agrees with the previous results \cite{AKS},
\begin{eqnarray}
J_w       & = & \frac{1}{6} (J+4 \D G) \mid, \;\;\;\;
J_s         =  \frac{1}{2} (J+2 \D G) \mid \; , \nn \\
G^{+}     & = & \frac{1}{\sqrt{2}} Q \mid, \;\;\;\;
G^{-}       =  -\sqrt{2} \D F \mid \; , \nn \\
T_w       & = & \frac{1}{(2c+3)} \left[ \frac{(1-2c)}{2} T-
\frac{(2c+1)}{2} [\D, \Db]
J+\frac{2}{3} J J+\frac{4}{3} J \D G \right. \nn \\
          &   & \left. +\frac{8}{3} \D G \D G-4 \D J G
          +\partial J \right] \mid \; , \nn \\
T_s       & = & \frac{1}{2} \left[ T+\partial J+2 \partial
\D G \right] \mid \; , \nn \\
S_1       & = & \frac{1}{\sqrt{2}} G \mid, \;\;\;\;
\bar{S}_1   =  -\frac{1}{\sqrt{2}} \D J \mid \; , \nn \\
S         & = & -F \mid, \;\;\;\;
\bar{S}     =  - \D Q \mid \; , \nn \\
S_2       & = & \frac{1}{\sqrt{2}} \left[-\frac{(2c+1)}{(2c-1)}
\Db J-\frac{(2c+3)}{
c(2c-1)} G \D G-\frac{(2c+1)}{c(2c-1)} J G+\partial G \right] \mid \; ,
\nn \\
\bar{S}_2 & = & \frac{1}{\sqrt{2}} \left[ \D T+\partial \D J-
\frac{1}{c} J \D J-
\frac{3}{c} \D J \D G \right] \mid \; ,
\label{eq:corr}
\end{eqnarray}
where $\mid$ means the $\theta, \bar{\theta}$ independent part of
corresponding composite operators. All four composite currents of
spins $(5/2,5/2,3,3)$ are given by
\bea
\Db F \mid,\; \Db T \mid,\; [\D, \Db] T \mid, \; [\D, \Db]F \mid\;.
\eea
By exploiting the constraint equations (\ref{eq:con2}),
these composite currents can be written through the basic twelve
elementary currents defined in (\ref{eq:corr}).

As was mentioned earlier, there is a trouble with checking Jacobi
identity for $ ( T, T, T ) $ directly in $N=2$ superspace. This
difficulty can be got round by the following argument. Using the
correspondence (\ref{eq:corr}) and resorting to the component
results of \cite{AKS} we are already guaranteed that the Jacobi
identities
\bea
( T, T, T ) \mid,\; ( T, T, \D T ) \mid,\; (
T, \D T, \D T ) \mid,\; ( \D T, \D T, \D T) \mid
\eea
are satisfied. Then other Jacobi identities containing
$ \Db T \mid, [ \D, \Db ] T \mid$
can be checked indirectly with making use of the above constraints,
taking into account that the l.h.s. of these identities can be written
as some composites involving $T \mid, \D T \mid$. The latter property
follows from the relations
\begin{eqnarray}
 \Db T\mid         =  -\frac{4}{(1 + 2c)} \left[ G T + Q F\right]\mid,
\end{eqnarray}
\begin{eqnarray}
 [ \D, \Db ] T\mid    = \left\{\partial T +  \frac{8}{(2c+1)} \left[ G \D T
 - Q \D F -\D G T - \D Q F \right] \right\} \mid \;.
\end{eqnarray}

One observes that there are extra composite currents in the r.h.s. of
(\ref{eq:alg}) which do not appear at the classical level
\cite{IKS}:
\begin{eqnarray}
F(Z_1)T(Z_2)_{q. corr.} & = &
\frac{\T \Tb}{\Z^2} \frac{16}{(2c+1)(2c-1)} G \D F \nn \\
T(Z_1)T(Z_2)_{q. corr.} & = &
\frac{\T \Tb}{\Z^2} \frac{16}{(2c+1)(2c-1)}
\left[G \D T-Q \D F \right]\;.
\end{eqnarray}
Only these three currents survive among more than $200$ ones we started
with. These surviving terms vanish in the appropriate classical limit,
$c \rightarrow \infty$, as it was discussed in \cite{AKS}. Also it can be
checked that after taking this limit all the structure constants present in
the SOPEs (\ref{eq:alg}) go into those appearing in the corresponding
SOPEs of \cite{IKS}.
\vspace{0.4cm}

\noindent{\bf 3. Conclusion}
\vspace{0.2cm}

\noindent To summarize, as a generalization and development of previous
findings [5 - 8] we have determined the full structure of
quantum $N=2$ super $W_3^{(2)}$ algebra in $N=2$ superspace. It would be
interesting to construct its free superfield realization which in
components reduces to the "hybrid" field realization found in \cite{AKS}.
\vspace{0.4cm}

\noindent{\bf Acknowledgements}
\vspace{0.2cm}

\noindent C. Ahn expresses his gratitude to the Directorate of
Bogoliubov Laboratory of Theoretical Physics, JINR, for hospitality
during the course of this work. The work of three of us
(E.I., S.K. \& A.S.) was supported in part by the RFFR grant
93-02-3821 and the INTAS grants 93-127 and 94-2317.

\end{document}